\documentclass{comnet}                     

\usepackage{graphicx}
\graphicspath{{Images/}}
\usepackage{stmaryrd}
\usepackage{color, soul}
\usepackage{amsmath, amsfonts, amssymb, amsthm}
\usepackage{afterpage}
\usepackage{float}
\usepackage{url}
\usepackage{ulem}
\usepackage{enumerate}

\bibpunct{[}{]}{,}{n}{,}{;}

\begin{document}

\title{Graph space: using both geometric and probabilistic structure to evaluate statistical graph models!}

\author{
\name{Louis Duvivier}
\address{Univ Lyon, INSA Lyon, CNRS, LIRIS UMR5205, F-69621 France. \email{louis.duvivier@insa-lyon.fr}} 
\and 
\name{R\'emy Cazabet}
\address{Univ Lyon, Universit\'e Lyon 1, CNRS, LIRIS UMR5205, F-69622 France.\email{remy.cazabet@univ-lyon1.fr}} 
\and 
\name{C\'eline Robardet}
\address{Univ Lyon, INSA Lyon, CNRS, LIRIS UMR5205, F-69621 France. \email{celine.robardet@insa-lyon.fr}}
}

\date{\today}

\maketitle

\begin{abstract}
{Statistical graph models aim at representing graphs as random realization among a set of possible graphs. To evaluate the quality of a model $M$ with respect to an observed network $G$, most statistical model selection methods rely on the probability that $G$ was generated by $M$, which is computed based on the entropy of the associated microcanonical ensemble. In this paper, we introduce another possible definition of the quality of fit of a model based on the edit distance expected value (EDEV). We show that adding a geometric structure to the microcanonical ensemble induces an alternative perspective which may lead to select models which could potentially generate more different graphs, but whose structure is closer to the observed network. Finally we introduce a statistical hypothesis testing methodology based on this distance to evaluate the relevance of a candidate model with respect to an observed graph.}
{statistical graph model, graph ensemble, entropy, statistical hypothesis testing, edit distance.}
\end{abstract}

\section{Introduction}
The study of large and complex systems in domains as diverse as physics, biology, computer science or social science brings forward interaction networks which are neither regular, nor random. The interaction structure follows no clear pattern, as in a grid for example, but it often presents remarkable properties such as a heterogeneous degree distribution (scale-free) \cite{barabasi1999emergence}, small average distances (the small-world phenomenon) \cite{watts1998collective}, a high level of transitivity and community structure \cite{newman2003structure}, which are not present in random graphs. This suggests that interactions are the result of a random process following some constraints which shape the resulting network. 

Various statistical models have been proposed to describe such constrained random graphs: the configuration model, which preserves degree distribution; the stochastic blockmodel, which preserves local densities; spatial models, which focus on distance between nodes \cite{simini2012universal}. A review of statistical graph models can be found in \cite{goldenberg2010survey} and an introduction to their formalism in \cite{cimini_statistical_2018}.

This diversity of parametric models raises the issue of model selection: considering one specific graph, which model is the most relevant? And with which set of parameters? Considering models as ensembles makes it possible to leverage results from statistical physics and information theory to perform model selection \cite{ding_model_2018}, in particular using maximum likelihood estimators \cite{stoica2004model} or equivalently the minimum description length \cite{grunwald_tutorial_2004}. In this framework, a model is considered to be good for a given graph if it generates it with high probability, \textit{i.e.} if the associated ensemble has a small entropy. Therefore, many papers have focused on computing the entropy of various graph ensemble \cite{bianconi_entropy_2009}, \cite{peixoto_entropy_2012}, \cite{zingg_what_2019}. A whole methodology for community detection based on those principles has been developed in the case of stochastic blockmodels \cite{peixoto_bayesian_2017}. 

However, the definition of the likelihood of a model based on the entropy of its microcanonical ensemble implies that by itself this likelihood cannot be interpreted, it can only be compared with the likelihood of another model for the same graph: It is fundamentally a relative measure. More generally, this definition of the likelihood makes it possible to identify the model that generates the observed graph with the highest probability, but it is not always the best possible measure of how relevant a model is with respect to an observed graph. Indeed, graphs are also geometrical objects, in the sense that one can define distances between them. Such a distance induces a structure on a model's ensemble. Much work has been devoted to quantifying how similar two graphs are \cite{wills_metrics_2019}, especially from a topological point of view \cite{monnig_resistance_2016}, \cite{koutra_deltacon:_2013}. These distances between two graphs can be generalized to evaluate the quality of a model by computing a distance between an observed graph and the graph ensemble associated to a model. For example, the widely used measure for community detection known as modularity \cite{newman2006modularity} evaluates the quality of a partition by comparing the edge weight in the observed graph with the expected edge weight of the graph in the configuration model, as developed in section \ref{sec_barycenter}. In this case, the problem is to evaluate the statistical significance of the results, in order not to mistake noise for structure \cite{guimera2004modularity}.

In this paper, after reviewing in section \ref{sec_entropy} and \ref{sec_barycenter} existing techniques to measure the relevance of a model with respect to a graph, we introduce in section \ref{sec_edev} the edit distance expected value, a measure which takes into account both the geometric and the probabilistic structure of the graph ensemble. Finally, we show how this measure can be used to evaluate a model relevance with respect to a given graph in section \ref{mod_likelihood}.

\section{Microcanonical ensemble}
The main point of statistical modeling is to describe the structure of a graph $G$ based on a set of global properties of that graph $P_M$ \cite{cimini_statistical_2018}. Some examples of common models and their associated properties are given in table \ref{mod_ex}. Apart from those properties, the graph is considered to be random. In practice, this means that $G$ is considered to have been chosen from the set:
\[\Omega_M = \{H \mid P_M(H) = P_M(G)\}\]
This set is called the microcanonical ensemble, in reference to statistical physics in which it was first introduced to represent all the possible states of a system corresponding to a global property. It can be defined with directed or undirected graphs, weighted or not, with or without self-loops. In the rest of the article, we will consider labelled directed multigraphs with self-loops. Although they are not the most widely used in practice, it makes probability derivations easier, especially for the configuration model.

\begin{table}[htb]
  \caption{Common statistical graph models and their associated properties.}
  \label{mod_ex}
  \begin{tabular}{|l|l|}
      \hline
    Model $M$ & Properties $P_M$ \\
    \hline
    Erd\"os-Renyi & number of nodes and number of edges \\
    Configuration model & degree distribution \\
    Stochastic block model & block to block density \\
    Gravity model & node position, strength, and deterrence function \\
    Radiation model & node position and strength \\
    \hline
  \end{tabular}
\end{table}

\subsection{Entropy}
\label{sec_entropy}
Information theory shows that, in order to ensure that no additional information bias the results, the probability distribution $\mathbb{P}_M$ on $\Omega_M$ has to be the one that maximises Shannon's entropy 
\[ S = -\sum_{H \in \Omega_M}\mathbb{P}(H)\mathrm{log}(\mathbb{P}(H)) \]
With no additional constraint, this optimal distribution is simply the uniform one:
\begin{equation}
    \label{uniform}
    \forall H \in \Omega_M, \mathbb{P}_M[H] = \frac{1}{|\Omega_M|}
\end{equation}
whose entropy is $\mathrm{log}(|\Omega_M|)$. Thus, computing the probability to choose $G$ among all possible graphs in $\Omega_M$ boils down to counting the number of graphs it contains. This has been done for different models in \cite{peixoto_entropy_2012}. 

The microcanonical ensemble and its entropy provide a common formulation for various statistical models. This is useful to perform model selection. Indeed, entropy is directly related to the likelihood of a given model. If we observe a graph $G$ and consider a set of models $\mathcal{M} = \{M_1, \dots, M_p\}$, we can find which model $G$ has most likely been sampled from by maximising its likelihood

\[ M^* = \underset{M_i \in \mathcal{M}}{\mathrm{argmax}} \; \mathbb{P}[M_i|G] \]

This maximisation can be done using Bayes theorem

\begin{equation}
    \label{bayes}
    \mathbb{P}[M_i|G] = \frac{\mathbb{P}[G|M_i] \times \mathbb{P}[M_i]}{\mathbb{P}[G]}
\end{equation} 

As $\mathbb{P}[G]$ does not depend on $M_i$, maximising the likelihood is equivalent to maximising the numerator of equation (\ref{bayes}). The first term of the product $\mathbb{P}[G|M_i]$ corresponds to the probability to generate $G$ with the model $M_i$, and according to equation (\ref{uniform}), maximising it is equivalent to minimising the entropy of the associated microcanonical ensemble $\log(|\Omega_{M_i}|)$. The second term $\mathbb{P}[M_i]$ is a prior distribution defined on the set of candidate models $\mathcal{M}$. Its role is to account for the fact that a model with enough parameters can be made arbitrarily close to any given graph $G$, up to the point where $\Omega_M = \{G\}$. Such a model would generate $G$ with probability $\mathbb{P}_M[G] = 1$, but it would be overfitting. The prior distribution prevents this from happening by assigning lower probabilities to models which have more parameters. Typical methods to counterweight models with too many parameters are the Akaike Information Criterion and Bayesian Information Criterion \cite{stoica2004model}. This idea was also developed and applied to the case of stochastic blockmodels in \cite{peixoto_bayesian_2017}.

All these methods rely on a definition of the likelihood of a model according to the uniform distribution over the microcanonical ensemble, and thus on the hypothesis that the most relevant model $M$ for a graph $G$ is the one which generates $G$ with the highest possible probability. It considers the microcanonical ensemble as a geometrically unstructured set, with no notion of distance between graphs, and thus it does not discriminate between two models $M_1$ and $M_2$ such that $|\Omega_{M1}| = |\Omega_{M_2}|$ but where $\Omega_{M_1}$ contains graph similar to $G$ while $M_2$ does not. This definition of the goodness of fit of a model is not always the best one. In particular, one can be interested in selecting the model that most probably generates graph similar to $G$, and not necessarily $G$ itself.

\begin{figure}[htb]
  \includegraphics[width=0.3\textwidth]{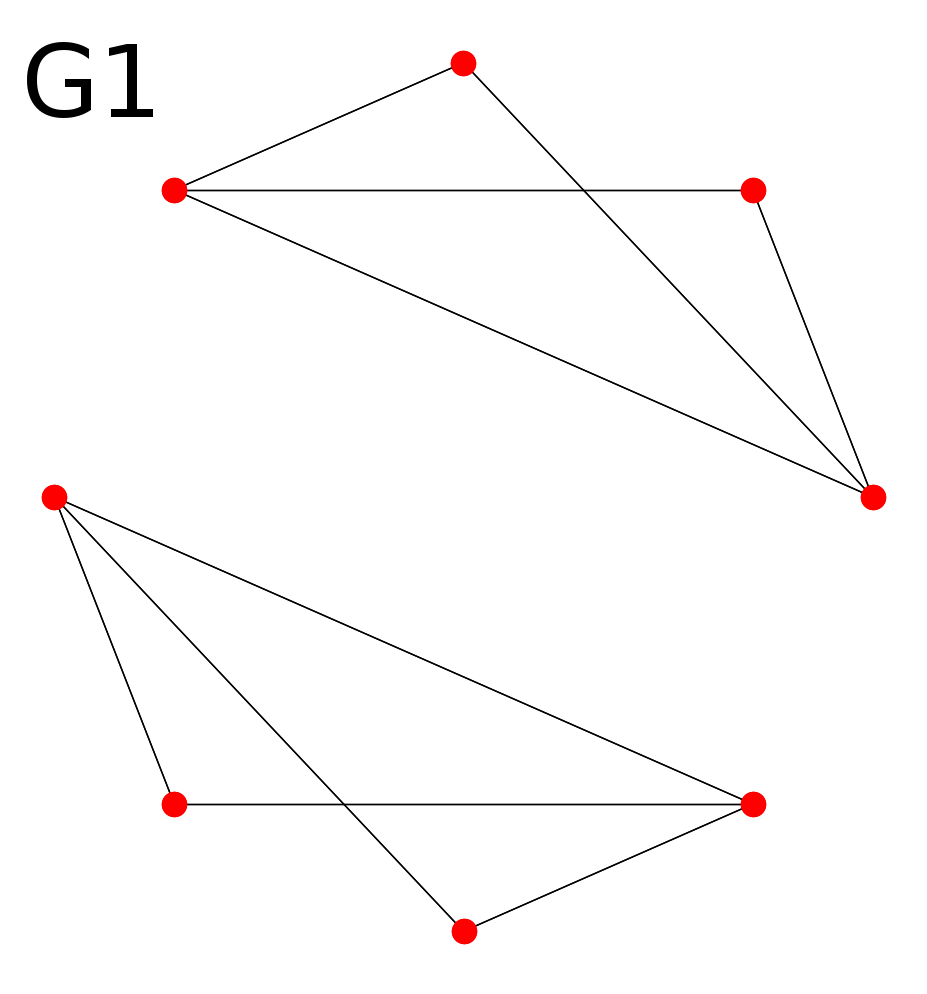}
  \hfill
  \includegraphics[width=0.3\textwidth]{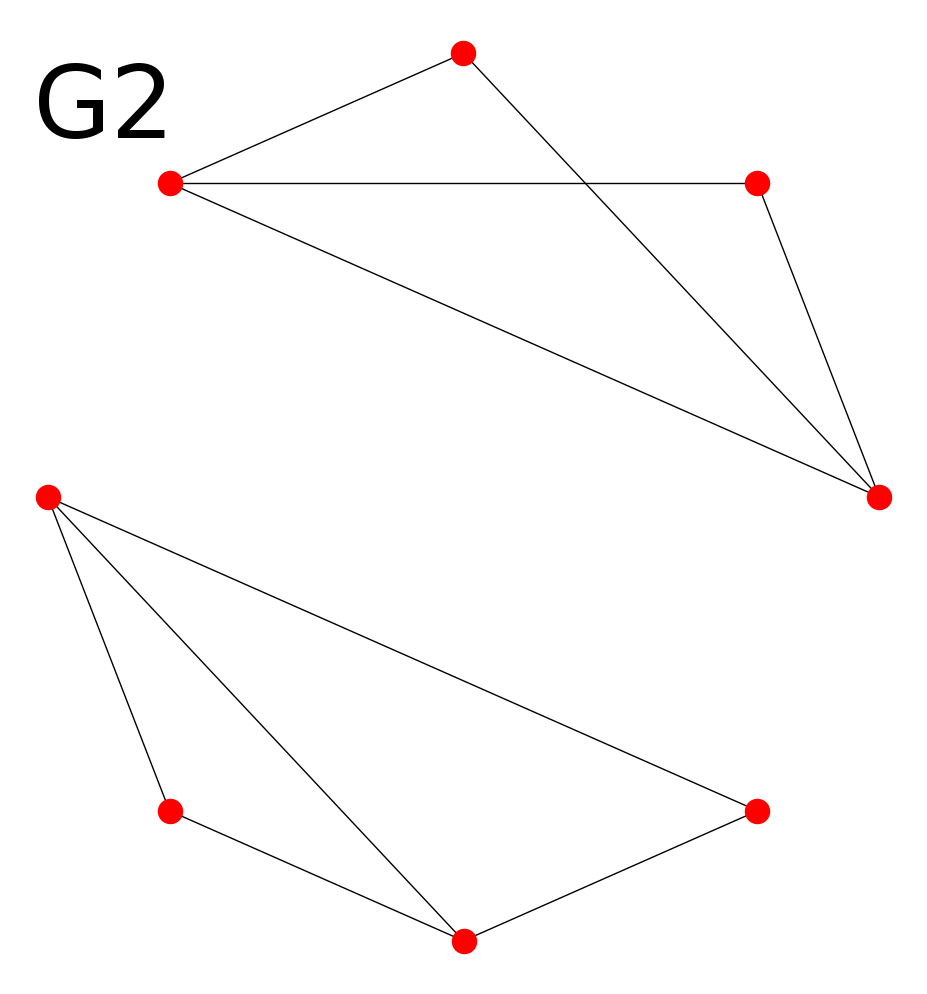}
  \hfill
  \includegraphics[width=0.3\textwidth]{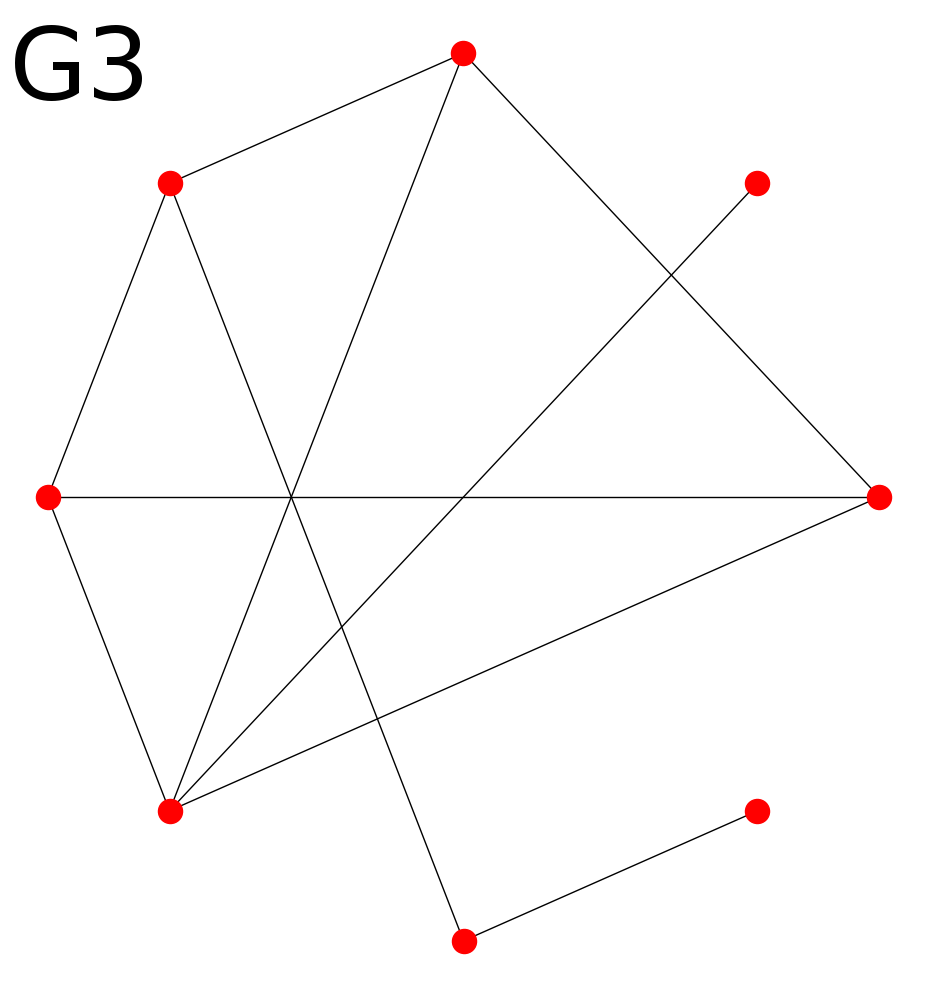}  
  \caption{\textbf{Three graphs with 8 nodes and 10 edges.} While $G_2$ is clearly more similar to $G_1$ than $G_3$, both $G_2$ and $G_3$ are different from $G_1$. Therefore, from the point of view of entropy maximisation, replacing $G_2$ by $G_3$ in the microcanonical ensemble does not affect the likelihood of a  model.\label{graphs_toy_example}}
\end{figure}

For example, if we consider the three graphs on figure \ref{graphs_toy_example} with $G_1$ as a reference, both $G_2$ and $G_3$ are different from $G_1$, but the topology of $G_1$ and $G_2$ is almost the same. Therefore, one could be interested in selecting a model which produces mostly $G_2$-like graphs rather than one which produces $G_3$-like ones. This cannot be done using the mere minimization of entropy since replacing $G_2$ by $G_3$ in the microcanonical ensemble does not change its entropy.

\subsection{Distance to the barycenter}
\label{sec_barycenter}
On the other hand, one may measure how typical $G$ is with respect to $\Omega_M$ by comparing it with an appropriate representative of this ensemble, for example its barycenter:
\[
  G_M = \sum_{H \in \Omega_M} \mathbb{P}(H) H
\]
If we denote $W_G$ the weight matrix of graph $G$, it can easily be derived that
\begin{equation}
  \label{barycenter}
  \forall (i,j) \in V^2, W_{G_M}(i,j) = \mathbb{E}[W_H(i,j)]
\end{equation}
\begin{remark}
$G_M$ does not necessarily belong to $\Omega_M$. In particular, even if all graphs in $\Omega_M$ have whole weight, it does not imply that $G_M$'s edge weights are integers. Examples of barycenter weights for common models are given in table \ref{barycenter_weight}

\begin{table}[htb]
  \caption{Statistical graph models' barycenter weight.}
  \label{barycenter_weight}
  \renewcommand{\arraystretch}{2}
  \begin{tabular}{|l|c|c|c|}
    \hline
    Model $M$ & Parameters & $G_M(i,j)$ & $m_{G_M}$\\
    \hline
    Erd\"os-Renyi & $n,m$ & $\frac{m}{n^2}$ & $m$ \\
    Configuration model & $(k_i^{in}, k_i^{out})_{i \in V}$ & $\frac{k_i^{out} k_j^{in}}{m}$ & $m$ \\
    Stochastic block model & $(m_{r,s})_{r,s}$ & $\frac{m_{c_i, c_j}}{|c_i||c_j|}$ & $m$ \\
    Gravity model & $(k_i)_{i \in V}, f$ & $k_i k_j f(d(i,j))$ & $m$ \\
    Radiation model & $(k_i^{in}, k_i^{out})_{i \in V}$ & $\frac{k_i^{out} k_i^{in} k_j^{in}}{(k_i^{in} + s_{ij})(k_i^{in} + k_j^{in} + s_{ij})} $ & $m$ \\
    \hline
  \end{tabular}
  \renewcommand{\arraystretch}{1}  
\end{table}
\end{remark}

The famous modularity function to evaluate the quality of a node partition $B = (b_1, \dots, b_p)$ on a graph $G = (V,E)$ with weight matrix $W_G$ is defined as the difference between the number of edges inside each cluster and the expected number for a random graph with the same degree distribution (\textit{i.e.} following the configuration model). It can be understood as a comparison with the barycenter $G_M$ of the corresponding configuration model.

\begin{align*}
  \mathrm{Q}(G, B) &= \frac{1}{2m} \sum_{i = 1}^p \sum_{u,v \in b_i} \left(W_G(u,v) - \frac{k_u^{out} k_v^{in}}{m} \right) \\
  &= \frac{1}{2m} \sum_{i = 1}^p \sum_{u,v \in b_i} \left(W_G(u,v) - W_{G_M}(i,j) \right) \\
  &= \mathrm{d}(G, G_M)
  \label{modularity}
\end{align*}

A problem is that $G$ is compared with a single graph $G_M$ which is supposed to account for the whole graph ensemble $\Omega_M$. In particular, all information about the dispersion around the barycenter is lost, which undermines any attempt to interpret statistically the results.

\section{Graph space and the edit distance expected value}
\label{sec_edev}
As we have seen, existing techniques to compare a graph $G$ and a model $M$ exploit in different ways the ensemble $\Omega_M$. Entropy based techniques described in section \ref{sec_entropy} focus on its cardinality, but they neglect the topological similarities of graphs inside the ensemble. On the other hand, as described in section \ref{sec_barycenter}, an objective function such as the modularity accounts for these similarities, but it does so with a single graph which is supposed to represent the whole set. Reality is more complex: $\Omega_M$ is a set of graphs with a probability distribution, and it can be further structured with a metric, making it a graph space. Both aspects, probabilistic and geometric, should be taken into account in order to understand the structure of $\Omega_M$, and the plausibility that a graph $G$ was generated by the associated model $M$.

Many different measures exist to compute a similarity score between two graphs $G$ and $H$ \cite{wills_metrics_2019}. One of the simplest is the edit distance. For two graphs on the same vertex sets $G_1 = (V, E_1)$ and $G_2 = (V, E_2)$, it counts the number of differences between their respective sets of edges.
\[ \mathrm{ed}(G_1, G_2) = \sum_{(i,j) \in V^2} |W_{G_1}(i,j) - W_{G_2}(i,j)|\]
As expected from its name, edit distance is a distance between graphs. Indeed, if we consider the weight matrix $W_G$ of a graph $G$ as a point in $\mathbb{R}^{n^2}$, the edit distance corresponds to the $\mathcal{L}_1$ distance and for any model $M$, $\Omega_M$ is a subset of $\mathbb{R}^{n^2}$. The dimension prevents any direct drawing of it for graphs with more than 2 nodes, but it is possible to obtain some intuition about its shape.

In the following, we will use a normalized version  of the edit distance which can be interpreted as the fraction of different edges between $G_1$ and $G_2$.
\begin{equation*}
  \mathrm{ned}(G_1, G_2) = \frac{1}{2m} \sum_{(i,j) \in V^2} |W_{G_1}(i,j) - W_{G_2}(i,j)|
\end{equation*}
This  normalized edit distance is no longer a distance on $\mathbb{R}^{n^2}$. Yet, for all models $M$ considered here, the number of edges $m$ is constant over the set $\Omega_M$. Thus, the normalized edit distance is equivalent to edit distance inside $\Omega_M$ and it allows to compare more easily results between various models, because whatever the model $M$, the distance between any two graphs $G_1$ and $G_2$ in $\Omega_M$ is at most $1$. 

\subsection{Edit distance to the barycenter}
\label{edit_dist_2_bar}

\begin{figure}[htb]
  \includegraphics[width=\textwidth]{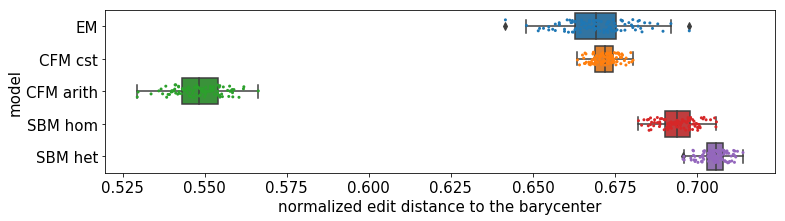}
  \caption{\textbf{Edit distance to the barycenter for 6 different models}. For each model $M$ in ordinate, we draw 100 graphs $G$ at random from $\Omega_M$ and compute for each of them $\mathrm{ned}(G, G_M)$. The distribution of results is then plotted as a boxplot. \label{ed_2_bar_dist}}
\end{figure}

$\Omega_M$ barycenter has already been introduced in section \ref{sec_barycenter}, where it was used as a proxy for the whole space. Using normalized edit distance, it is possible to check how much graphs in $\Omega_M$ are similar to the barycenter $G_M$. We consider six different models:
\begin{enumerate}
    \item EM: Erd\"os-Renyi with 50 nodes and 1000 edges
    \item CFM cst: configuration model with 50 nodes and a constant degree distribution ($k_{i}^{in} = k_i^{out} = 20$)
    \item CFM arith: configuration model with 50 nodes and an arithmetic degree distribution ($k_i^{in} = k_i^{out} = i + 1$)
    \item SBM hom: stochastic block model with 50 nodes and 5 communities, each having internal density 1.2, and external density 0.2.
    \item SBM het: stochastic block model with 50 nodes and 5 communities, with internal density 0.4, 0.8, 1.2, 1.6, 2, and external density 0.2.
\end{enumerate}
For each model $M$, we pick a random sample $\mathcal{S}_M$ of $100$ graphs in $\Omega_M$ and for all $G \in \mathcal{S}_M$ we compute the normalized edit distance to the barycenter $\mathrm{ned}(G, G_M)$. Results are shown in figure \ref{ed_2_bar_dist}.

The first thing to underline is that whatever the model, $\mathrm{ned}(G, G_M)$ is greater than 0.5, which means that most graphs in $\Omega_M$ have at most half of their edges in common with $G_M$. This observation shows that for those models, the graph space is not concentrated around its barycenter. On the contrary, most graphs in $\Omega_M$ seem to be at a specific distance from its barycenter, as would happen for a sphere with a radius depending on the model: 0.67 for ER and CFM cst, 0.55 for CFM arith, 0.69 for SBM hom and 0.71 for SBM het.

All models were chosen to have similar entropy, as shown in table \ref{model_entropy}, yet their characteristic distance to the barycenter vary greatly. Furthermore, we observe that these quantities are not positively correlated: CFM arith, which is the model with the larger entropy is also the one which is the most concentrated around its barycenter. This means that even if this model can generate a higher number of different graphs, the graphs it produces tend to be more similar one to the other than for other models. This is logical as this model preserves a degree distribution, which enforces more constraints on edges' distribution than an Erd\"os-Renyi or a stochastic block model. 

\begin{table}[htb]
\begin{center}

    \begin{tabular}{c|c|c}
            Model & Characteristic distance & Entropy \\
            \hline
            ER & 0.67 & 2050 \\
            CFM cst & 0.67 & 2500 \\
            CFM arith & 0.55 & 3300 \\
            SBM hom & 0.69 & 1840 \\
            SBM het & 0.71 & 1840 \\
        \end{tabular}
        \caption{\textbf{Edit distance to the barycenter and entropy.}\label{model_entropy}}
        \end{center}
\end{table}

This concentration of graphs at a specific distance from the barycenter is a consequence of the dimensionality of the vector space. Let's denote $\mathcal{B}(G, r)$ the ball of center $G$ and radius $r$ in $(\mathbb{R}^{n^2}, \mathrm{ed})$. We consider the set 
\begin{align*}
    \Omega_M(r) &= \{G \in \Omega_M |\, \mathrm{ned}(G, G_M) \leq r\} \\
    &= \{G \in \Omega_M |\, \mathrm{ed}(G, G_M) \leq 2mr\} \\
    &= \Omega_M \cap \mathcal{B}(G_M, 2mr)
\end{align*}
The volume $V_n(r)$ of $\mathcal{B}(G_M, 2mr)$ is proportional to $r^{n^2}$, therefore 
\begin{equation}
    \label{ball_volume}
    \forall r < 1, \frac{V_n(r)}{V_n(1)} \underset{n \rightarrow \infty}{\longrightarrow} 0
\end{equation}
The volume of the ball concentrates quickly at its periphery as the dimension increases, and so does the volume of $\Omega_M$. The additional constraints on $\Omega_M$ modify its shape in such a way that graphs too far away from the barycenter are rare, which explains why the concentration does not happen at distance $1$ from the barycenter. Still, this phenomenon is strong enough to imply that even graphs generated according to a model $M$ will share only a relatively small fraction of their edges with the barycenter of the model.

\subsection{Edit distance expected value}

The previous observations on the structure of graph spaces show that in order to compare a graph $G$ with a model $M$, one should consider more than the mere cardinal of $\Omega_M$. One possibility to evaluate how similar to $G$ are the graphs in $\Omega_M$ is to compute the expected value of the normalized edit distance:

\[
  \mathrm{EDEV}(G, M) = \underset{H \in \Omega_M}{\mathbb{E}} \left [\frac{1}{2m} \sum_{(i,j) \in V^2} |W_{G}(i,j) - W_{H}(i,j)|\right ] 
  \label{edit_dist_expected_value}
\]

\begin{figure}[htb]
  \begin{center}
  \includegraphics[height=0.25\textheight]{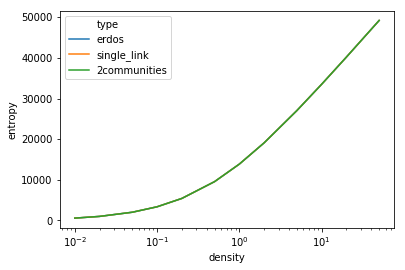}\\
  \includegraphics[height=0.25\textheight]{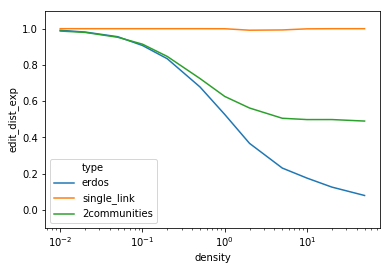}
  \end{center}
  \caption{\textbf{Entropy and edit distance expected value against density.} For each density, three graphs are generated. $G_1(m)$ is random, $G_2(m)$ is made of two random communities and $G_3(m)$ has its edges concentrated on a single pair of nodes of weight $m$. On the top plot, the entropy $\mathrm{log}(|\Omega_{ER(n,m)}|)$ is plotted against the density $\frac{m}{n^2}$. As all graphs belong to the same graph ensemble $\Omega_{ER(n,m)}$, the three curves are the same. On the bottom plot the edit distance expected value $EDEV(G(m), \Omega_{ER(n,m)})$ is plotted against density.
  \label{ent_edev_vs_dens}}
\end{figure}

To illustrate how EDEV provides further information on the place of $G$ within the graph space, we compare it with entropy for different synthetic graphs. A low value indicates that $G$ is close to other graphs in $\Omega_M$, and thus that it is typical of the model, while a high value shows that it is an outlier. As a case study, we consider the Erd\"os-Renyi model. Let's recall that we consider multigraphs, which implies that we allow for densities rising above $1$. The extension of Erd\"os-Renyi model to multigraphs is straightforward, $\Omega_{ER(n,m)}$ contains all multigraphs with $n$ nodes and $m$ edges and each multigraph is generated with the same probability $\frac{1}{|\Omega_{ER(n,m)}|}$. In practice, we consider models with $n = 100$ nodes and a number of edges $m$ ranging from $100$ to $500000$. For each, we consider three graphs:
\begin{itemize}
\item $G_1(m)$, picked uniformly at random inside $\Omega_{\mathrm{ER}(n,m)}$
\item $G_2(m)$, a graph made of two equal communities, each with $\frac{n}{2}$ nodes and $\frac{m}{2}$ edges, perfectly separated.
\item $G_3(m)$, the graph where all edges are between nodes 0 and 1.
\end{itemize}
Results are shown on figure \ref{ent_edev_vs_dens}.

For each value of $m$, all three graphs belong to the graph ensemble $\Omega_{\mathrm{ER}(n,m)}$. We observe that as density increases $|\Omega_{ER(n,m)}|$ grows exponentially, which implies that the probability to pick at random $G_1(m)$, $G_2(m)$ or $G_3(m)$ becomes even less probable. Yet, in the case of the random graph $G_1(m)$ this is counter-intuitive: as density grows and becomes higher than 1, most graph in $\Omega_{\mathrm{ER}(n,m)}$ become complete graphs with each edge having weight about $\frac{m}{n^2}$. This is the case of $G_1$ too with a high probability, so $\mathrm{ER}(n,m)$ is very likely to produce graphs similar to $G_1(m)$, even if it is very unlikely to produce $G_1(m)$ itself.

On the other hand, edit distance expected value is able to capture this phenomenon. While it is close to 1 for all three types of graphs when density is low because in this situation a random model can hardly predict correctly which edge is present in any graph, it decreases quickly towards 0 when density rises above 0.1 for $G_1(m)$. For $G_2(m)$ we have an intermediate situation: edit distance decreases too, but it reaches its minimum around 0.5, indicating that even when it is densely populated, the model is only able to reproduce correctly half of its edges. This is normal as $G_2(m)$ concentrates them inside the communities, which means on half of all possible node pairs.
These observations are actually a particular case of a more general result, which can be stated as:

\begin{lemma}
    \label{lemma_1}
    Let B be a partition of $\llbracket 1, n\rrbracket$ with $p$ blocks. Let $M \in \mathcal{M}_{p}(\mathbb{N})$ be a block adjacency matrix. For all $k \in \mathbb{N}$, we define the stochastic blockmodel $S(k) = (B, k \cdot M)$, and its barycenter $G_{S(k)}$. We consider a sequence of random graphs $(G_k)_{k \in \mathbb{N}}$, each drawn from $S(k)$. We have that
    
    \begin{equation*}
        \mathrm{ed}(G_k, G_{S(k)}) \overset{\mathbb{P}}{\underset{k \rightarrow \infty}{\longrightarrow}} 0
    \end{equation*}
\end{lemma}

and as a consequence:

\begin{theorem}
Let $B_1$ and $B_2$ be two partition on $\llbracket 1, n \rrbracket$, with $p_1$ and $p_2$ blocks respectively. Let $M_1 \in \mathcal{M}_{p_1}(\mathbb{N})$ and $M_2 \in \mathcal{M}_{p_2}(\mathbb{N})$ be two block adjacency matrices such that 
\[ \sum_{i,j \in [1, p_1]^2} M_1[i,j] = \sum_{k,l \in [1, p_2]^2} M_2[k,l] = m \]
Let's consider two series of stochastic blockmodels defined as $S_1(k) = (B_1, k \cdot M_1)$ and $S_2(k) = (B_2, k \cdot M_2)$, whose barycenters are denoted $G_1(k)$ and $G_2(k)$. We have that

\begin{enumerate}
    \item There exists $d \in \mathbb{R}, \forall k \in \mathbb{N}, \mathrm{ed}(G_1(k), G_2(k)) = d$
    \item Let $(G_k)_{k \in \mathbb{N}}$ be a series of random graphs, each drawn following model $S_1(k)$. 
\end{enumerate}
\begin{equation}
    \mathrm{EDEV}(G_k, S_2(k)) \overset{\mathbb{P}}{\underset{k \rightarrow \infty}{\longrightarrow}} d
\end{equation}
(proof in Appendix)
\end{theorem}

\begin{remark}
In particular, if $M_1 = M_2$, lemma \ref{lemma_1} means that the normalized edit distance expected value between a graph picked at random and the barycenter of the stochastic blockmodel converges toward $0$: $\Omega_{S(k)}$ shrinks around $G_{S(k)}$. This is what we observe with $G_1(m)$. Yet, we also observe on figure \ref{ent_edev_vs_dens} that the normalized edit distance converges toward $0$ only as density rises above $1$. Thus, in practice, the vast majority of graphs are too sparse for this assumption to hold true and most graphs in $\Omega_{S(k)}$ are far from $G_{S(k)}$, as developed in section \ref{edit_dist_2_bar}
\end{remark}


\section{Statistical test}
\label{mod_likelihood}
As the distance to the barycenter, the expected value of the normalized edit distance is characteristic of a model. For a model $M$, the values of $\mathrm{EDEV}(H,M)$ for graphs $H$ in $\Omega_M$ are concentrated around a specific value $d_M$. We can use this fact to rule out models which fit badly on an observed graph $G$.

\begin{figure}
    \centering
    \includegraphics[width=\textwidth]{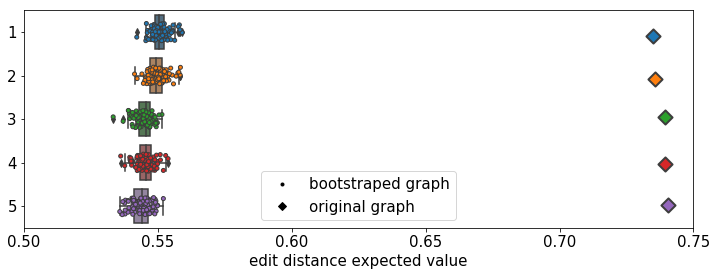}
    \caption{$5$ graphs $G_i$ were generated with a generative model $CFMD(n, k_i^{out}, k_i^{in})$. Their normalized edit distance expected value with respect to a candidate model $SBM_i$ is plotted as diamond. As a comparison point, the distribution of the normalized edit distance expected value for $100$ test graphs generated with $SBM_i$ is plotted as dots and boxplot.}
    \label{edev_dist_gcfmd_msbm}
\end{figure}

For example, let's consider the configuration model $CFMD(n, k_i^{out}, k_i^{in})$ 
\begin{gather*}
    n=50 \\
    \forall i \in \llbracket 0, n-1 \rrbracket, k_i^{out} = k_i^{in} = i
\end{gather*}
We use this model to generate a graph $G_i$. $CFMD(n, k_i^{out}, k_i^{in})$ will thus be called the \textit{generative} model, and $G_i$ the \textit{observed} graph. On this observed graph, we test the stochastic blockmodel $SBM_i$ obtained by partitioning its nodes in two blocks: $B_0$ contains even nodes and $B_1$ odd nodes (this way we avoid to put all high-degree nodes in the same block) and learning the block adjacency matrix on $G_i$. We call $SBM_i$ the \textit{candidate} model. We generate a sample $\mathcal{S}_i$ of $100$ test graphs with the candidate model $SBM_i$ and compare the normalized edit distance $\mathrm{EDEV}(G_i, \Omega_{SBM_i})$ of the observed graph to the candidate model with $\mathrm{EDEV}(H, \Omega_{SBM_i})$ for all test graphs $H \in \mathcal{S}_i$. This experiment is performed $5$ times, and results are shown on figure \ref{edev_dist_gcfmd_msbm}.

\begin{figure}
    \centering
    \includegraphics[width=\textwidth]{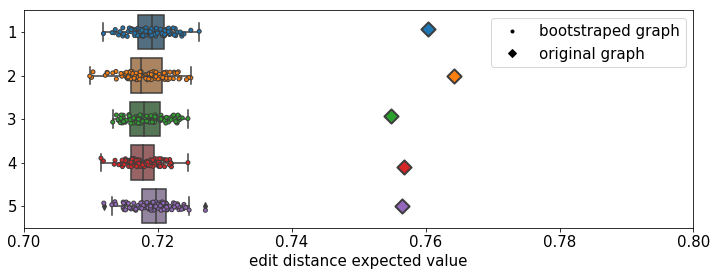}
    \caption{$5$ graphs $G_i$ were generated with a generative model $SBM(n, B, M)$. Their normalized edit distance expected value with respect to a test model $CFMD_i$ is plotted as diamond. As a comparison point, the distribution of the normalized edit distance expected value for $100$ test graphs generated with $CFMD_i$ is plotted as dots and boxplot.}
    \label{edev_dist_gsbm_mcfmd}
\end{figure}

\begin{figure}
    \centering
    \includegraphics[width=\textwidth]{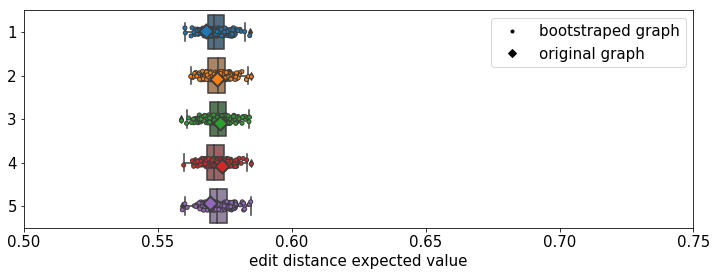}
    \caption{$5$ graphs $G_i$ were generated with a generative model $SBM(n, B, M)$ which is also used as the test model. Their normalized edit distance expected value with respect to the test model is plotted as diamond. As a comparison point, the distribution of the normalized edit distance expected value for $200$ other graphs generated with $SBM(n,B,M)$ is plotted as dots and boxplot.}
    \label{edev_dist_gsbm_msbm}
\end{figure}

We observe that for all five experiments, the normalized edit distance expected value to the candidate model $SBM_i$ for the observed graph $G_i$ is around $0.74$, while for the test graphs generated by $SBM_i$ it is between $0.53$ and $0.56$. This shows that the normalized edit distance expected value to $SBM_i$ is significantly different for the observed graphs, which are generated by $CFMD(n, k_i^{out}, k_i^{in})$, and for the test graphs generated by $SBM_i$. It is thus very unlikely that the observed graph $G_i$ was generated by the candidate model $SBM_i$.

We then perform the same experiment the other way round, by considering as generative model a stochastic blockmodel $SBM_0(n, B, M)$ defined by
\begin{gather*}
    n = 50\\
    B = \llbracket 0, 24 \rrbracket, \llbracket 25, 49\rrbracket \\
    M = \begin{bmatrix}
    500 & 0 \\
    0 & 500
    \end{bmatrix}
\end{gather*}
5 graphs $G_i$ are generated using this stochastic blockmodel. As candidate model, we consider a configuration model $CFMD_i$ obtained by learning the degree sequence of $G_i$. A sample $\mathcal{S}'$ of $100$ test graphs is randomly picked in $\Omega_{CMFD_i}$ and we compare $\mathrm{EDEV}(G_i, \Omega_{CFMD_i})$ with $\mathrm{EDEV}(H, \Omega_{CFMD_i})$ for all $H \in \mathcal{S}'$. Results are shown on figure \ref{edev_dist_gsbm_mcfmd}. Once again, we observe that the normalized edit distance expected value to the candidate model $CFMD_i$ is significantly different for the observed graphs, which were generated by $SBM(n, B, M)$, and for the test graphs generated by $CFMD_i$. This allows us to reject the hypothesis that the observed graph $G_i$ was generated by the candidate model $CFMD_i$.

Finally, as a null case, we consider the situation in which the generative model and the test model are the same. We generate $5$ graphs $(G_i)_{i \in \llbracket 1, 5 \rrbracket}$ with the same stochastic blockmodel $SBM_0(n, B, M)$ as before, and consider this same model $SBM_0$ as the candidate model. A sample $\mathcal{S}''$ of $200$ test graphs is randomly picked in $\Omega_{SBM_0}$ and we compare $\mathrm{EDEV}(G_i, \Omega_{SBM_0})$ with $\mathrm{EDEV}(H, \Omega_{SBM_0})$ for all $H \in \mathcal{S}''$. Results are shown on figure \ref{edev_dist_gsbm_msbm}. In this case, we observe that the normalized edit distance expected value for the first $5$ graphs is not significantly different from the one of the sampled graphs. We cannot reject the hypothesis that the observed graph $G_i$ was generated by $SBM_0$.

\afterpage{\clearpage}

\subsection{Statistical hypothesis testing}
The methodology can be formalized using statistical hypothesis testing. Let's say we have a graph $G$ and a model $M$ (possibly obtained by fitting some parameters on $G$). We want to test the null hypothesis $\mathcal{H}_0$: ``The observed graph $G$ has been generated by the candidate model $M$". If we knew the distribution of the edit distance expected value to the model $M$ for graphs generated by $M$, we could simply compute $\mathrm{EDEV}(G,M)$ and evaluate the probability to obtain a value at least as large under the hypothesis that $G$ was generated by $M$. However, characterizing the distribution $\mathbb{P}_{EDEV}(M)$ for various models and deriving the parameters of the distribution from the parametrization of the model would require to further investigate the structure of their associated microcanonical ensembles.

As we do not know this distribution, we use bootstrapping to conduct our test. This approach consists in replacing the unknown distribution by a sample generated using this distribution. We therefore use $M$ to generate a sample of $q$ graphs $G_1, \dots G_q$. We can then consider $G$ as a sample of size $1$ from an unknown distribution $M'$ and $\mathcal{H}_O$ can be reformulated as: "$(G_i)_{i \in \llbracket 1, q \rrbracket}$ and $G$ were generated by the same probability distribution". Such an hypothesis can be tested using Fisher's permutation test, as described in \cite{efron1994introduction}. We consider for each graph its edit distance expected value to the model $M$: $x_i = \mathrm{EDEV}(G_i, M)$ and $y= \mathrm{EDEV}(G,M)$, and compute the difference between the means of the two samples 
\[ \theta = \bar{x} - \bar{y} = \frac{1}{q} \sum_{i=1}^q x_i - y \]
We test $\mathcal{H}_0$ by evaluating, under this hypothesis, the probability $p = \mathbb{P}_{\mathcal{H}_0}[\theta^* \geq \theta]$ to obtain a value of $\theta^*$ at least as large for two samples of size $1$ and $q$ generated by $M$. The exact probability cannot be computed, as we do not know the probability distribution $\mathbb{P}_{EDEV}(M)$, but it can be approximated by considering all the pairs of samples of size $1$ and $q$ that can be constructed by picking at random one graph within the $q+1$ graphs $G, G_1, \dots, G_q$. We denote $\mathcal{P}_j$ the pair of samples $\{(G_j), (G_1, \dots, G_{j-1}, G, G_{j+1}, \dots, G_q)\}$. For this pair of samples, the difference between the means is denoted 
\[ \theta^*_j = \frac{1}{q} \left( \sum_{i \neq j} x_i + y \right) - x_j \]
If $\mathcal{H}$ is true, the $q+1$ pairs of samples $\mathcal{P}_i$ are a subset of the possible pairs of samples generated by $M$, and we can make the approximation
\[ \mathbb{P}_{\mathcal{H}_0}[\theta^* \geq \theta] \approx \frac{\#\{j \mid \theta^*_j \geq \theta\}}{q+1} \]
If $\mathbb{P}_{\mathcal{H}_0}[\theta^* \geq \theta]$ is lower than a preset threshold $\delta$ (in the following we will use $\delta = 0.01$), one can conclude that there is enough evidence to reject the hypothesis that $G$ was generated by $M$ (and thus affirm that $G$ was not generated by $M$). On the other hand, if it is greater than this threshold, there is not enough evidence to reject it. 
It should be stressed that this last sentence does not mean that one can affirm that $G$ was generated by $M$, but only that this hypothesis cannot be discarded.

\begin{figure}[htb]
    \begin{center}
    \includegraphics[scale=0.5]{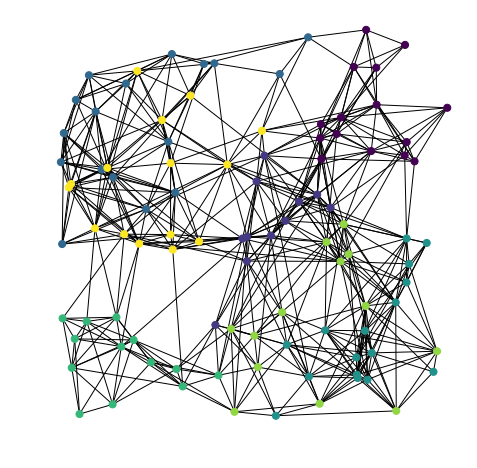}
    \end{center}
    \caption{\label{spatial_graph} Spatial graph generated using the Waxman random geometric model. $100$ nodes are randomly distributed in a $[0,1]\times[0,1]$ square. They are then connected with a probability depending on the distance $d$ between nodes: $p(d) = 10\exp\left(\frac{d}{0.05L}\right)$, with $L$ the maximum distance between two nodes. Communities are computed using graph-tool
    (\protect\url{https://graph-tool.skewed.de/}).}
\end{figure}

Let's consider a situation in which one wishes to evaluate the relevance of the block structure computed on a graph $G$, with a weight matrix $W_G$. Whatever the graph, and the partition $B = (b_1, \dots, b_p)$ of its $n$ nodes, it is always possible to define $M \in \mathcal{M}_p(\mathbb{N})$ as
\[ \forall i,j \in \llbracket 1, p \rrbracket, M[i,j] = \sum_{u \in b_i, v \in b_j} W_G[u,v]\]
such that $G \in \Omega_{SBM(n, B, M)}$. The objective is to evaluate whether this stochastic blockmodel is a relevant model of $G$. An even trickier question is to evaluate whether any stochastic blockmodel can be a relevant model. In particular, spatial models can generate graphs with groups of nodes densely connected due to their position rather than to block membership. An example of such a graph is shown in figure \ref{spatial_graph}. It may then be hard to tell whether the blocks found are indeed a legitimate model of the observed graph or should be considered as artefacts, consequences of the underlying spatial structure.

To illustrate how statistical hypothesis testing allows to address this issue, we consider eight models: four stochastic blockmodels and four Waxman models for random geometric graphs\footnote{\url{https://networkx.org/documentation/stable/reference/generated/networkx.generators.geometric.waxman\_graph.html}} with different sets of parameters. The Waxman model for spatial graphs allows to easily control the strength of the spatial structure, by tuning the speed at which edge probability decays as the distance between nodes rises. The number of nodes is fixed to $n = 100$ and the parameters are fixed such as to ensure a density $d$ around $0.036$. All stochastic blockmodels use a node partition in four blocks of $25$ nodes, with a block adjacency matrix of the form:
\[
\begin{bmatrix}
    m_{int} & m_{ext} & m_{ext} & m_{ext} \\
    m_{ext} & m_{int} & m_{ext} & m_{ext} \\
    m_{ext} & m_{ext} & m_{int} & m_{ext} \\
    m_{ext} & m_{ext} & m_{ext} & m_{int}
\end{bmatrix}
\]  
The four stochastic blockmodels are then defined by:
\begin{enumerate}
    \item $M_0$: $m_{int} = 90$, $m_{ext} = 0$    
    \item $M_1$: $m_{int} = 75$, $m_{ext} = 5$
    \item $M_2$: $m_{int} = 60$, $m_{ext} = 10$
    \item $M_3$: $m_{int} = 45$, $m_{ext} = 15$
\end{enumerate}
This way, the graphs generated using $SBM_0$ are made of perfectly separated blocks of nodes, while those generated by $SBM_3$ have blocks with as many internal and external edges.

For the Waxman models, we also consider four parameter sets:
\begin{enumerate}
    \item $M_4$: $\alpha = 0.1$, $\beta = 1$
    \item $M_5$: $\alpha = 0.08$, $\beta = 1.6$
    \item $M_6$: $\alpha = 0.06$, $\beta = 2.7$
    \item $M_7$: $\alpha = 0.04$, $\beta = 8.5$
\end{enumerate}
The lower the value of $\alpha$, the stronger the spatial structure.

\begin{figure}
    \centering
    \includegraphics[scale=0.35]{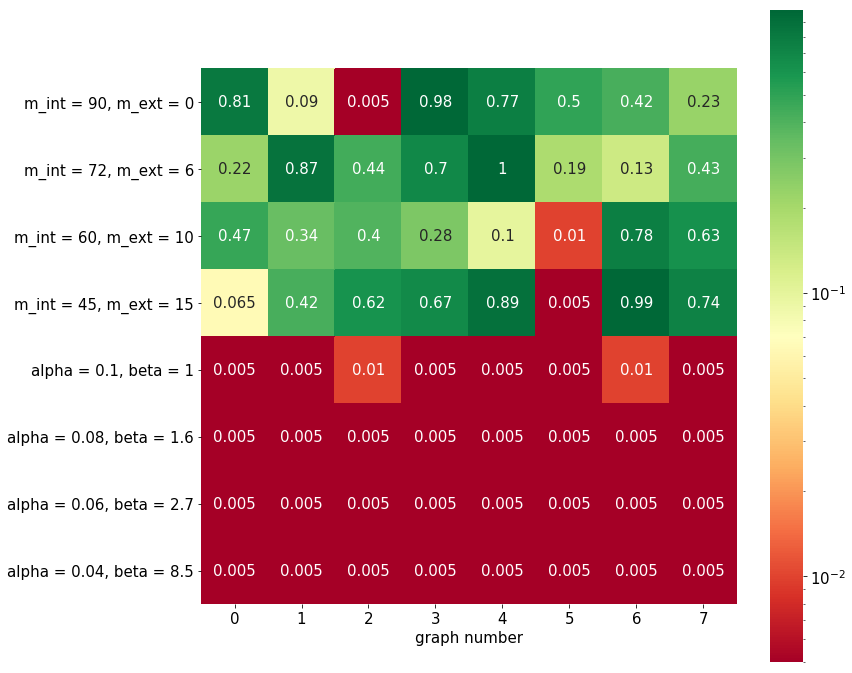}
    \caption{For each model $M_i$ in ordinate, and each graph number $i$ in abscissa, we compute the probability $p_{i,j}$ that a graph generated using $SBM_{i,j}$ has a normalized edit distance expected value to $SBM_{i,j}$ further away from the mean value $d_{i,j}$ than $G_{i,j}$. The probability is plotted in heatmap.}
    \label{p_value}
\end{figure}

With each model $M_i$, we generate $8$ graphs $(G_{i,j})_{j \in \llbracket 0, 7 \rrbracket}$. On each of those observed graph, we find the minimum entropy node partition $B_{i,j}$ using graph tools, and fit a candidate stochastic blockmodel $SBM_{i,j}$ on $G_{i,j}$ based on this node partition. We then evaluate the relevance of this stochastic blockmodel using the previously described methodology. We use a confidence level $\delta$ of $0.01$ and a sample size $q = 200$. The probabilities $p_{i,j} = \mathbb{P}_{\mathcal{H}_0}[\theta^*_{i,j} \geq \theta_{i,j}]$ obtained are plotted in figure \ref{p_value}.

We observe that for all graphs generated by stochastic blockmodels but three, $p_{i,j} > 10^{-2} = \delta$. On the other hand, for all spatial graphs $p_{i,j} \leq \delta$. This means that the hypothesis "$G_{i,j}$ has been generated by the candidate stochastic block model $SBM_{i,j}$" is rejected for all $32$ spatial graphs, and for three of the stochastic blockmodels generated graphs. Let's stress again that a probability $p_{i,j}$ superior to $\delta$ does not mean that $SBM_{i,j}$ is the right model for $G_{i,j}$. It only means that there is not enough statistical evidence to reject it.

These results show that, on $32$ spatial graphs generated with various sets of parameters, the statistical hypothesis testing methodology is able to correctly identify that the block structure found is not a relevant model, despite being the solution maximizing the likelihood among the candidate blockmodels. This result is not trivial. Indeed, the fact that this partition is the most likely according to the entropy definition of likelihood means that the block identified are characterized by a specific connection density which cannot be explained by random fluctuations. However, as the likelihood is maximized on a set of models which does not contain the Waxman model used to generate the graph, this estimator is not able to recognize that its structure would be even better explained using a spatial model. What is more, this methodology manages to reject the block structure for all spatial graphs while spuriously rejecting it for only three out of $32$ graphs originally generated with a stochastic blockmodel. In other words, there is no false positive, and only three false negatives.

Strictly speaking, these results only allow to rule out one node partition. As the block structure tested were fitted on the observed graphs using minimum entropy, one could argue that ruling out this partition implies that no other node partition can lead to a relevant model. Yet, for most real graphs, there is more than one plausible node partition and minimizing the entropy of a partition is a stochastic process. Therefore, the experiment should be performed more than once to conclude that the observed graph has no block structure.

\section{Conclusion}
As a conclusion, we have seen that widely used quality measures for graph models rely either on the number of different graphs they can produce, which neglects the geometric structure of the graph space, or on a direct comparison with the barycenter of those graphs, which discards information about the distribution around this barycenter. Because of these restrictions, they are unable to distinguish between graphs which have a typical structure of a model and graphs which may be generated by this model but as outliers. 

In this paper, we show how edit distance can provide additional information on the structure of the graph space which is captured neither by the entropy nor the barycenter. By computing the expected value of the normalized edit distance for a given graph, we obtain a criterion which can be used to evaluate the model quality with respect to this graph. Finally, we incorporate this criterion to a statistical hypothesis testing methodology to perform model selection. 

Graph space is a theoretical framework which can be used for any statistical model, and particularly spatial models. It allows to compare them with SBM or configuration model, and perform model selection between models of different nature. What is more, statistical hypothesis testing provides a statistically rigorous methodology to evaluate the relevance of a candidate model to an observed graph. 


Apart from its simplicity to compute and interpret, an interesting result about the normalized edit distance expected value is that the same quantity can be used to test many different models. Statistical tests could be performed on any graph property, such as the average path length or the clustering coefficient, to rule out a candidate model. Yet, such tests require to choose for each model and each tested graph a property that could be used as a statistical test. In turn, all these properties can be used to define new models which better reproduce the values measured on the observed graph (as done for example when defining degree-corrected stochastic blockmodels, which incorporate the degree distribution constraint to the block structure). The risk being that by taking more and more properties into account, one eventually overfits the graph. 
Note, however, that the expected value of the normalized edit distance is somewhat different from the previously mentioned properties. Indeed, as it depends on both the graph and the whole microcanonical ensemble, it is impossible to compute its value on a graph observed beforehand and then to define a model adapted to this specific value.

Indeed, the fact that, for most models, the normalized edit distance expected value distribution quickly clusters around the mean, is a property of the graph space rather than the graph itself. It is a consequence of a geometrical result (the volume of a ball in $n$ dimensions), which highlights the benefits of considering the geometric structure of graph ensembles. However, the edit distance is not the only distance that can be used. Considering other metrics which are more sensitive to the global topology of the network, like the perturbation-resistance distance or spectral distances, could provide additional insight on the structure of the graph space.

\clearpage
\section{Appendix}

\subsection{Barycenter graph weight of various statistical models}
We have defined the barycenter of a graph model as
\[
  G_M = \sum_{H \in \Omega_M} \mathbb{P}(H) \cdot H
\]
Which means that
\begin{align*}
  \forall (i,j) \in V^2, W_{G_M}(i,j) &= \sum_{H \in \Omega_M} \mathbb{P}(H) \times W_H(i,j) \\ &=  \mathbb{E}[W_H(i,j)]
\end{align*}
Let's illustrate how this can be computed for some classical models.

\paragraph{Erd\"os-R\'enyi model} The simplest graph model is the Erd\"os-R\'enyi model for random graphs. It's associated microcanonical ensemble can be defined as:
\[ \Omega_{ER(n,m)} = \{H = (V,E) \mid |V| = n \land |E| = m\}\]
Let's recall that for the sake of simplicity, we chose to consider multigraphs with self loops. Thus, the computation of $\mathbb{E}[W_H(i,j)]$ is particularly simple. Indeed, if for each pair of node $(i,j) \in V^2$ and each $k \in [1,m]$ we define the random variable $X_{i,j,k}$ which is equal to $1$ if the $k^{th}$ edge is $i \rightarrow j$ and to $0$ else, then we have that $W_H(i,j) = \sum_{k = 1}^m X_{i,j,k}$. It is a sum of independent Bernouillis' random variable so it follows a binomial law of parameters $m$ and $\frac{1}{n^2}$, and thus
\begin{equation}
    W_{G_{ER(n,m)}}(i,j) = \underset{\Omega_{ER}}{\mathbb{E}}[W_H(i,j)] = \frac{m}{n^2}
\end{equation}

\paragraph{Configuration Model} For the configuration model, all graphs in the microcanonical ensemble must have the same degree distribution. Let's consider the directed version. 
\[ \Omega_{CFMD} = \{G \mid \forall i \in V,  \mathrm{deg}_G^{out}(i) = k_i^{out} \land \mathrm{deg}^{in}_G(i) = k_i^{in}\}\]
To compute the weight of the barycenter graph's edges, we consider that each node $i$ has $k_i^{out}$ outgoing stubs and $k_i^{in}$ ingoing stubs. Any graph in $\Omega_{CFMD}$ is characterized by a configuration of connections of outgoing stubs with ingoing stubs. For every pair of nodes $i,j \in V^2$ and any pair of stub $k \in [1, k_i^{out}]$, $l \in [1, k_j^{in}]$, we define the random variable $X_{i,j,k,l}$ which is equal to $1$ if the $k^{th}$ outgoing stub of $i$ is connected to the $l^{th}$ ingoing stub of $j$, and to $0$ otherwise. Then
\[ W_{H}(i,j) = \sum_{k = 1}^{k_i^{out}} \sum_{l = 1}^{k_j^{in}} X_{i,j,k,l} \]
As each outgoing stub of $i$ has the same probability to be connected to any of the $m$ ingoing stubs
\[ \forall i,j,k,l, \mathbb{P}[X_{i,j,k,l} = 1] = \frac{1}{m}\]
Thus, $W_H(i,j)$ follows a binomial law of parameters $\frac{1}{m}$ and $k_i^{out} \times k_j^{in}$. Finally
\begin{equation}
W_{G_{CFMD}}(i,j) = \underset{\Omega_{CFMD}}{\mathbb{E}}[W_H(i,j)] = \frac{k_i^{out} \times k_j^{in}}{m}
\end{equation}

\paragraph{Stochastic blockmodel} The case of the stochastic blockmodel can be treated in the same way as erd\"os-r\'enyi. It is defined, considering a partition of the nodes $B = (b_1, \dots, b_q)$ and a block adjacency matrix $M \in \mathcal{M}_q(\mathbb{N})$ by
\[ \Omega_{SBM} = \left\{H \mid \forall b_k,b_l, \sum_{i \in b_k} \sum_{j \in b_l} W_H(i,j) = M(k,l) \right\}\]
So, for any pair of nodes $i \in b_k, j \in b_l$, $W_H(i,j)$ follows a binomial law of parameters ($M(k,l)$, $|b_k||b_l|$). Thus
\begin{equation}
W_{G_{SBM}}(i,j) = \underset{\Omega_{SBM}}{\mathbb{E}}[W_H(i,j)] = \frac{M(k,l)}{|b_k||b_l|}
\end{equation}

\paragraph{Spatial models} References for the gravitational model and the radiation model can be found in \cite{barthelemy2011spatial} and \cite{simini2012universal}. In both cases, they are constructed in such a way that edges weight have a given expected value. In the case of the gravitational model, it is
\begin{equation}
W_{G_{\mathrm{grav}}}(i,j) = \mathrm{f}(\mathrm{d}(i,j)) \times k_i^{out} \times k_j^{in}
\end{equation}
where $\mathrm{d}(i,j)$ is the distance from node $i$ to node $j$, and $\mathrm{f}$ is a deterence function.

Finally, in the case of the radiation model, it is
\begin{equation}
    W_{G_{\mathrm{rad}}}(i,j) = \frac{k_i^{out} \times k_i^{in} \times k_j^{in}}{(k_i^{in} + s_{ij}) \times (k_i^{in} + k_j^{in} + s_{ij})}
\end{equation}
with $s_{ij} = \sum_{u \in \mathcal{C}(i,j)} k_u^{in}$ and $\mathcal{C}(i,j) = \{u \in V \mid 0 < \mathrm{d}(i,u) < \mathrm{d}(i,j)\}$.

\clearpage
\subsection{Convergence proof}
First of all, let's prove the following lemma

\begin{lemma}
    \label{lemma}
    Let B be a partition of $\llbracket 1, n\rrbracket$ with $p$ blocks. Let $M \in \mathcal{M}_{p}(\mathbb{N})$ be a block adjacency matrix. For all $k \in \mathbb{N}$, we define the stochastic blockmodel $S(k) = (B, k \cdot M)$, and its barycenter $G_{S(k)}$. We consider a sequence of random graphs $(G_k)_{k \in \mathbb{N}}$, each drawn from $S(k)$. We have that
    
    \begin{equation*}
        \mathrm{ed}(G_k, G_{S(k)}) \overset{\mathbb{P}}{\underset{k \rightarrow \infty}{\longrightarrow}} 0
    \end{equation*}
\end{lemma}

Given the notation above, we want to prove that:
\begin{equation*}
    \forall \alpha > 0, \mathbb{P}[\mathrm{ed}(G_k, G_{S(k)}) > \alpha] \underset{k \rightarrow \infty}{\longrightarrow} 0
\end{equation*}

Let $\alpha > 0$. Let's denote $m = \sum_{i, j} M_{i,j}$ the number of edges of graphs in $\Omega_{S(1)}$. By definition, 
\[ \mathrm{ed}(G_k, G_{S(k)}) = \frac{1}{2km} \sum_{u,v} |W_{G_k}(u,v) - W_{G_{S(k)}}(u,v)|\]
Thus, 
\begin{equation*}
    \mathrm{ed}(G_k, G_{S(k)}) > \alpha \Rightarrow \exists (u,v), \left|\frac{W_{G_k}(u,v) - W_{G_{S(k)}}(u,v)}{2km}\right| > \frac{\alpha}{n^2}
\end{equation*}
and
\begin{equation*}
    \mathbb{P}[\mathrm{ed}(G_k, G_{S(k)}) > \alpha] \leq \sum_{u,v} \mathbb{P}\left[ \left|\frac{W_{G_k}(u,v) - W_{G_{S(k)}}(u,v)}{2km}\right| > \frac{\alpha}{n^2} \right]
\end{equation*}

Let's consider two blocks $b_i$ and $b_j$ in $B$. We know that $\forall u \in b_i,v \in b_j, W_{G_{S(k)}}(u,v) = k \cdot \frac{M_{i,j}}{|b_i||b_j|}$ and $W_{G_k}(u,v) \sim \mathcal{B}(k \cdot M_{i,j}, p_{i,j})$ with $p_{i,j} = \frac{1}{|b_i||b_j|}$. Therefore, according to the Bienaym\'e-Tchebychev inequality:

\begin{align*}
    \mathbb{P}\left[\left|\frac{W_{G_k}(u,v) - W_{G_{S(k)}}}{2km}\right| > \frac{\alpha}{n^2} \right] &\leq \frac{k \times M_{i,j} \times p_{i,j} \times (1-p_{i,j}) \times n^2}{4 \times k^2 \times m^2 \times \alpha} \\
    &\leq \frac{M_{i,j} \times p_{i,j} \times (1-p_{i,j}) \times n^2}{4 \times k \times m^2 \times \alpha} \\
    &\underset{k \rightarrow \infty}{\longrightarrow} 0
\end{align*}

Thus, 

\begin{equation}
    \mathbb{P}[\mathrm{ed}(G_k, G_{S(k)}) > \alpha] \underset{k \rightarrow \infty}{\longrightarrow} 0
\end{equation}
Which proves the lemma.

\newpage
We can now prove the theorem

\begin{theorem}
Let $B_1$ and $B_2$ be two partition on $\llbracket 1, n \rrbracket$, with $p_1$ and $p_2$ blocks respectively. Let $M_1 \in \mathcal{M}_{p_1}(\mathbb{N})$ and $M_2 \in \mathcal{M}_{p_2}(\mathbb{N})$ be two block adjacency matrices such that 
\[ \sum_{i,j \in [1, p_1]^2} M_1[i,j] = \sum_{k,l \in [1, p_2]^2} M_2[k,l] = m \]
Let's consider two series of stochastic blockmodels defined as $S_1(k) = (B_1, k \cdot M_1)$ and $S_2(k) = (B_2, k \cdot M_2)$, whose barycenters are denoted $G_1(k)$ and $G_2(k)$. We have that

\begin{enumerate}
    \item There exists $d \in \mathbb{R}, \forall k \in \mathbb{N}, \mathrm{ed}(G_1(k), G_2(k)) = d$
    \item Let $(G_k)_{k \in \mathbb{N}}$ be series of random graph each drawn following model $S_1(k)$. 
\end{enumerate}
\begin{equation*}
    \mathrm{EDEV}(G_k, S_2(k)) \overset{\mathbb{P}}{\underset{k \rightarrow \infty}{\longrightarrow}} d
\end{equation*}
\end{theorem}

For any pair of nodes $i,j$, belonging to blocks $b(i)$ and $b(j)$ in $B_1$ (resp. $B_2$), the weight of the edge $i \rightarrow j$ in $G_1(k)$ (resp. $G_2(k)$) is given by:
\[ W_{G_1(k)}[i,j] = k \cdot \frac{M[b(i), b(j)]}{|b(i)||b(j)|} \]

Therefore, the edit distance between $G_1(k)$ and $G_2(k)$ is 
\begin{align*}
    \mathrm{ed}(G_1(k), G_2(k)) &= \frac{1}{2km} \sum_{i,j \in [1,n]^2} \left|  W_{G_1(k)}[i,j] - W_{G_2(k)}[i,j]\right| \\
    &= \frac{1}{2km} \sum_{i,j \in [1,n]^2} \left|  k \cdot \frac{M_1[b_1(i), b_1(j)]}{|b_1(i)||b_1(j)|} - k \cdot \frac{M_2[b_2(i), b_2(j)]}{|b_2(i)||b_2(j)|}\right| \\
    &= \frac{1}{2m} \sum_{i,j \in [1,n]^2} \left|\frac{M_1[b_1(i), b_1(j)]}{|b_1(i)||b_1(j)|} - \frac{M_2[b_2(i), b_2(j)]}{|b_2(i)||b_2(j)|}\right|
\end{align*}
which is constant with respect to $k$. In the following we will denote this distance $d$ for the sake of conciseness. We want to show that
\[ 
\mathrm{EDEV}(G_k, S_2(k)) \overset{\mathbb{P}}{\underset{k \rightarrow \infty}{\longrightarrow}} d 
\]

We start by noticing that
\begin{align*}
    \mathrm{EDEV}(G_k, S_2(k)) - d &= \underset{H \in S_2(k)}{\mathbb{E}}[\mathrm{ed}(G_k, H)] - d\\
    & \leq \underset{H \in S_2(k)}{\mathbb{E}}[\mathrm{ed}(G_k, G_1(k)) + \mathrm{ed}(G_1(k), G_2(k)) + \mathrm{ed}(G_2(k), H)] - d\\
    & \leq \mathrm{ed}(G_k, G_1(k)) + \underset{H \in S_2(k)}{\mathbb{E}}[\mathrm{ed}(G_2(k), H)]
\end{align*}

On the other hand, 
\begin{align*}
    d - \mathrm{EDEV}(G_k, S_2(k)) &= \underset{H \in S_2(k)}{\mathbb{E}}[\mathrm{ed}(G_1(k), G_2(k)) - \mathrm{ed}(G_k, H)] \\
    &\leq \underset{H \in S_2(k)}{\mathbb{E}}[\mathrm{ed}(G_1(k), G_k) + \mathrm{ed}(G_k, H) + \mathrm{ed}(H, G_2(k)) - \mathrm{ed}(G_k, H)] \\
    &\leq \mathrm{ed}(G_k, G_1(k)) + \underset{H \in S_2(k)}{\mathbb{E}}[\mathrm{ed}(G_2(k), H)] \\
\end{align*}

Thus
\begin{equation}
    \label{fond_ineq}
    |\mathrm{EDEV}(G_k, S_2(k)) - d| \leq \mathrm{ed}(G_k, G_1(k)) + \underset{H \in \Omega_{S_2(k)}}{\mathbb{E}}[\mathrm{ed}(G_2(k), H)]
\end{equation}

Because $G_k$ is generated following $S_1(k)$, a direct application of lemma \ref{lemma} is that 
\[ \mathrm{ed}(G_k, G_1(k)) \overset{\mathbb{P}}{\underset{k \rightarrow \infty}{\longrightarrow}} 0\]
What is more, if $H$ is generated following $S_2(k)$, we also have that 
\[\mathrm{ed}(H, G_2(k)) \overset{\mathbb{P}}{\underset{k \rightarrow \infty}{\longrightarrow}} 0\]
which implies that $\mathrm{ed}(H, G_2(k)) \overset{\mathcal{L}}{\underset{k \rightarrow \infty}{\longrightarrow}} 0$ and in particular 
\[\underset{H \in \Omega_{S_2(k)}}{\mathbb{E}}[H, G_2(k)] \underset{k \rightarrow \infty}{\longrightarrow} 0\]

Finally, we obtain that 
\[ \mathrm{ed}(G_k, G_1(k)) + \underset{H \in \Omega_{S_2(k)}}{\mathbb{E}}[H, G_2(k)] \overset{\mathbb{P}}{\underset{k \rightarrow \infty}{\longrightarrow}} 0\]

And thanks to equation \ref{fond_ineq}:
\begin{equation}
    \mathrm{EDEV}(G_k, S_2(k)) \overset{\mathbb{P}}{\underset{k \rightarrow \infty}{\longrightarrow}} d
\end{equation}


\section*{Funding}
Work supported by the ACADEMICS grant of the IDEXLYON, project of the Université de Lyon, PIA operated by ANR-16-IDEX-0005 and the BITUNAM grand ANR-18-CE23-0004.

\end{document}